\documentclass[letter]{aa} 

\usepackage{graphicx}
\usepackage{txfonts}
\usepackage{amsmath}    
\usepackage{amssymb}    
\usepackage[hidelinks]{hyperref}
\usepackage{xcolor}
\hypersetup{colorlinks, linkcolor={blue!70!black}, citecolor={blue!70!black}, urlcolor={blue!70!black}
}
\def\hii{\relax \ifmmode {\mbox H\,\textsc{ii}}\else H\,{\scshape ii}\fi}
\def\hi{\relax \ifmmode {\mbox H\,\textsc{i}}\else H\,{\scshape i}\fi}

\begin{document} 

\title{The Large Magellanic Cloud stellar content with SMASH:}

\subtitle{I. Assessing the stability of the Magellanic spiral arms}

\titlerunning{Assessing the stability of the Magellanic spiral arms}

\author{
T. Ruiz-Lara\inst{1, 2}, C. Gallart\inst{1, 2}, M. Monelli\inst{1, 2}, D. Nidever\inst{3, 4}, A. Dorta\inst{1, 2}, Y. Choi\inst{3, 5}, K. Olsen\inst{4}, G. Besla\inst{6}, \\ E.J. Bernard\inst{7}, S. Cassisi\inst{8, 9}, P. Massana\inst{10}, N.E.D. No\"el\inst{10}, I. P\'erez\inst{11, 12}, V. Rusakov\inst{13}, M.-R. L. Cioni\inst{14},  S. R. Majewski\inst{15}, R.P. van der Marel\inst{5, 16}, D. Mart\'inez-Delgado\inst{17}, A. Monachesi\inst{18, 19}, L. Monteagudo\inst{20, 1, 2}, R.R. Mu\~noz\inst{21}, \\ G.S. Stringfellow\inst{22}, F. Surot\inst{1,2}, A. K. Vivas\inst{23}, A.R. Walker\inst{23}, \and D. Zaritsky\inst{5}
}

\authorrunning{Ruiz-Lara et al.}

\institute{
Instituto de Astrof\'isica de Canarias, Calle V\'ia L\'actea s/n, E-38205 La Laguna, Tenerife, Spain \\ \email{tomasruizlara@gmail.com}
\and
Departamento de Astrof\'isica, Universidad de La Laguna, E-38200 La Laguna, Tenerife, Spain
\and
Department of Physics, Montana State University, P.O. Box 173840, Bozeman, MT 59717-3840
\and
NSF's National Optical-Infrared Astronomy Research Laboratory, 950 North Cherry Ave, Tucson, AZ 85719
\and
Space Telescope Science Institute, 3700 San Martin Drive, Baltimore, MD 21218
\and
Steward Observatory, University of Arizona, 933 North Cherry Avenue, Tucson, AZ 85721, USA
\and
Universit\'e C\^ote d'Azur, OCA, CNRS, Lagrange, France
\and
INAF -- Astronomical Observatory of Abruzzo, via M. Maggini, sn, 64100 Teramo, Italy
\and
INFN, Sezione di Pisa, Largo Pontecorvo 3, 56127 Pisa, Italy
\and
Department of Physics, University of Surrey, Guildford, GU2 7XH, UK
\and
Departamento de F\'isica Te\'orica y del Cosmos, Universidad de Granada, Campus de Fuentenueva, E-18071 Granada, Spain
\and
Instituto Carlos I de F\'isica Te\'orica y computacional, Universidad de Granada, E-18071 Granada, Spain
\and
Department of Physics and Astronomy, University College London, Gower Street, WC1E 6BT London, UK
\and
Leibniz-Institut f\"{u}r Astrophysik Potsdam (AIP), An der Sternwarte 16, D-14482 Potsdam Germany
\and
Department of Astronomy, University of Virginia, Charlottesville, VA 22904-4325, USA
\and
Center for Astrophysical Sciences, Department of Physics \& Astronomy, Johns Hopkins University, Baltimore, MD 21218, USA
\and
Instituto de Astrof\'isica de Andaluc\'ia, CSIC, E-18080, Granada, Spain
\and
Instituto de Investigaci\'on Multidisciplinar en Ciencia y Tecnolog\'ia, Universidad de La Serena, Ra\'ul Bitr\'an 1305, La Serena, Chile
\and
Departamento de Astronom\'ia, Universidad de La Serena, Av. Juan Cisternas 1200 Norte, La Serena, Chile
\and
Isaac Newton Group of Telescopes, Apartado 321, 38700 Santa Cruz de La Palma, Canary Islands, Spain
\and
Departamento de Astronom\'ia, Universidad de Chile, Camino del Observatorio 1515, Las Condes, Santiago, Chile
\and
Center for Astrophysics and Space Astronomy, University of Colorado, 389 UCB, Boulder, CO, 80309-0389, USA
\and
Cerro Tololo Inter-American Observatory, NSF's National Optical-Infrared Astronomy Research Laboratory, Casilla 603, La Serena, Chile
}

   \date{Received --, 2020; accepted --, 2020}

 
  \abstract
   {The Large Magellanic Cloud (LMC) is the closest and most studied example of an irregular galaxy. Among its principal defining morphological features, its off-centred bar and single spiral arm stand out, defining a whole family of galaxies known as the Magellanic spirals (Sm). These structures are thought to be triggered by tidal interactions and possibly maintained via gas accretion. However, it is still unknown whether they are long-lived stable structures. In this work, by combining photometry that reaches down to the oldest main sequence turn-off in the colour-magnitude diagrams (CMD, up to a distance of $\sim$4.4 kpc from the LMC centre) from the SMASH survey and CMD fitting techniques, we find compelling evidence supporting the long-term stability of the LMC spiral arm, dating the origin of this structure to more than 2~Gyr ago. The evidence suggests that the close encounter between the LMC and the Small Magellanic Cloud (SMC) that produced the gaseous Magellanic Stream and its Leading Arm (LA) also triggered the formation of the LMC's spiral arm. Given the mass difference between the Clouds and the notable consequences of this interaction, we can speculate that this should have been one of their closest encounters. These results set important constraints on the timing of LMC-SMC collisions, as well as on the physics behind star formation induced by tidal encounters.
}

   \keywords{methods: observational -- techniques: photometric -- galaxies: stellar content -- galaxies: evolution -- galaxies: Magellanic Clouds}

   \maketitle
%

\section{Introduction}
\label{intro}

Magellanic spiral galaxies (Sm) are ubiquitous in the Universe. Characterised by the presence of an off-centred bar (SBm) and one single arm, S(B)m galaxies are some of the most structurally lopsided galaxies ever discovered \citep[][]{1972VA.....14..163D}. How they became so asymmetric, particularly with regard to the ultimate origin of their spiral arm(s) in contrast to more massive spiral galaxies, remains unclear. Fortunately, the level of detail attainable for the Large Magellanic Cloud (LMC), the prototype for this class of galaxy and the only example for which individual stars can be studied, could help us answer these open questions. 

Tidal interactions and minor mergers have long been invoked to explain asymmetries in galaxies  \citep[e.g.][]{1994AJ....107.1320O, 1997ApJ...477..118Z, 2002ApJ...580..705K, 2016ApJ...825...20B, 2018ApJ...866...90C, 2018ApJ...869..125C}. Theoretical studies conclude that brief tidal interactions can transform barred late-type spirals to SBm-like discs \citep[e.g.][]{2014MNRAS.439.1948Y}, and single-armed structures have been found in numerical studies of galaxy mergers \citep[e.g.][]{2003MNRAS.341..343B, 2018MNRAS.480.3069P}. Using $N$-body simulations, \citet[][]{2016ApJ...827..149P} found that dwarf-dwarf encounters mainly affect the disc of SBm galaxies, causing a mismatch between the centre of the bar and the dynamical centre of the galaxy. For the particular case of the LMC, Small Magellanic Cloud, and Milky Way (LMC-SMC-MW) system, \citet[][]{2012MNRAS.421.2109B} developed $N$-body smoothed-particle hydrodynamics modelling of interactions between the three galaxies that could simultaneously explain the observed large-scale features (the gaseous Magellanic Bridge, Stream, and Leading Arm), together with the LMC's off-centred bar and single arm. The authors claim that LMC-SMC interactions are the main driver of the observed morphology and configuration, with little effect from the MW. Indeed, the oddity of triple systems such as the LMC-SMC-MW in both observation \citep[e.g.][]{2011ApJ...733...62L} and simulation \citep[e.g.][]{2010MNRAS.406..896B}, in contrast with the large number of S(B)ms in the Universe \citep[][]{2009IAUS..256..461W, 2013ApJ...772..135Z}, suggests that Sm characteristics can be acquired without the presence of a nearby MW-like galaxy.

However, the high incidence of S(B)m and lopsided galaxies in the Local Universe and their presence in isolation \citep[][]{2004AJ....127.1900W, 2005A&A...438..507B, 2011A&A...530A..30V} suggests that asymmetries have to be long-lived --- if acquired through interaction-related processes --- or that other mechanisms might be at play. Cosmological gas accretion has been proposed as one of these complementary processes \citep[][]{2005A&A...438..507B}. If such accretion is asymmetric, it can induce lopsided and one-armed discs in the stellar component once the gas is converted into stars. Other studies of more massive systems, point to misalignments between discs and the galaxy's dark matter halo, regardless of their origin, as drivers of asymmetric discs \citep[][]{1998ApJ...496L..13L, 2001MNRAS.328.1064N}. Interestingly, several studies invoking different mechanisms predict a wide range of lifetimes for the observed asymmetries.

Weak interactions are found to cause short-lived asymmetries ($\sim$0.5~Gyr) and trigger star formation in the discs of lopsided galaxies \citep[][]{2000ApJ...538..569R}. \citet[][]{2014MNRAS.439.1948Y} also claim that asymmetries driven in LMC-mass galaxies via brief tidal interactions are short-lived (of the order of a Gyr or less), although they can shape long-lived structures in lower mass systems. Slightly more stable asymmetries seem to be induced in the case of dwarf-dwarf close interactions: \citet[][]{2016ApJ...827..149P} find that asymmetries can persist for $\sim$2~Gyr, until the disc is re-centred with the bar. \citet[][]{2016ApJ...825...20B} argue that one-armed structures can persist 1-2~Gyr after the smaller companion has been completely consumed, undergoing  reformation if several close encounters take place. Nevertheless, it seems that the combination of tidal interaction and mergers with cosmological accretion of gas can extend the life of asymmetries by several Gyrs \citep[][]{2005A&A...438..507B}. \citet[][]{1998ApJ...496L..13L} and \citet[][]{2001MNRAS.328.1064N} also showed that once an offset arises between discs and dynamical centres, they tend to remain offset. 

Within this context, it is clear that the precise dating of morphological structures in the LMC will provide unprecedented constraints to numerical modelling of SBm galaxies. In this work, we use photometric data from the Survey of the MAgellanic Stellar History \citep[SMASH,][]{2017AJ....154..199N} and colour-magnitude diagram (CMD) fitting to characterise the star formation history (SFH) of the outer-disc region of the LMC, with special focus on its spiral arm. We find compelling evidence on the stability of this morphological structure. 

\section{Determination of the LMC star formation history}
\label{analysis}

A detailed description of our process to study the LMC SFH using the SMASH data will be provided in a subsequent paper (Ruiz-Lara et al. in prep.). For completeness, we give a self-contained but brief explanation here. We analysed a total of 5.5$\times$10$^7$ stars from the main body of the LMC (up to a galactocentric distance of $\sim$4.4~kpc or 5$^{\circ}$) observed by the SMASH survey\footnote{SMASH DR2 (D. Nidever et al. in prep.) is available through the Data Lab hosted by the NSF's National Optical-Infrared Astronomy Research Laboratory.}. SMASH provides high-quality and deep photometry of the observed region in an uniform and homogeneous way\footnote{The SMASH data reduction is extensively described in \citet[][]{2017AJ....154..199N} and makes use, among other packages, of the PHOTRED (\citealt[][]{2011ApJ...733L..10N}, \url{https://github.com/dnidever/PHOTRED}) software package which performs multi-exposure forced-PSF photometry using the DAOPHOT suite of programs \citep[][]{1987PASP...99..191S, 1994PASP..106..250S}}, enabling the study of the SFH of this system as never before. The analysed sample of stars has been drawn from the original SMASH dataset after some quality cuts are applied (${\rm -2.5}$~<~{\it SHARP$_{DAOPHOT, ALLSTAR}$}~<~2.5; photometric errors in $g$ and $i$-bands below 0.3 mag). In this process, the catalogues of overlapping fields are combined and apparent magnitudes transformed to absolute magnitudes (on a star-by-star basis) taking into account reddening, distance, and LMC inclination effects (changing the effective distance modulus of different regions due to the orientation of the LMC disc) following the analysis presented in \citet[][]{2018ApJ...866...90C}. The particular choice of this geometry has little effect on our results and conclusions.

We split this initial catalogue into 232 individual spatial bins following a Voronoi tesselation \citep[][]{2003MNRAS.342..345C}. The goal is to divide the whole LMC field into homogeneous, compact, roundish, and contiguous spatial bins avoiding holes or overlaps, each containing $\sim$200,000 stars from the above-described sample. The photometric completeness and uncertainties in each spatial bin (critical for subsequent steps) have been derived following standard procedures of artificial-star tests \citep[ASTs; e.g.][]{2010ApJ...720.1225M}, injecting $\sim$2$\times$10$^6$ stars per spatial bin covering the range of colours, magnitudes, and sky locations sampled by the analysed stars. We assign to each of the spatial bins its corresponding AST table by cross-matching their HEALPix unique identifiers with those of the AST catalogues. 

A synthetic population containing 1.5$\times$10$^8$ stars was then created with uniform distributions in age and metallicity (from 0.03 to 14 Gyr and Z from 0.0001 to 0.025) based on the solar-scaled BaSTI stellar evolution models \citep[][]{2004ApJ...612..168P}. We use a Kroupa initial mass function \citep[][]{2001MNRAS.322..231K} and a binary fraction of 50$\%$ with a minimum mass ratio of 0.1. Observational uncertainties and crowding effects are simulated in this synthetic population using {\tt DisPar} \citep[see][and Ruiz-Lara et al. in prep. for more information]{2020arXiv200209714R} and the previously described AST tables. This enabled us to create a separate `dispersed' synthetic CMD that accounts for the crowding and photometric uncertainties that are characteristic of each of the 232 individual spatial bins. 

Finally, SFHs, defined as the star formation rate (SFR) and chemical enrichment as a function of cosmic time, are computed by CMD fitting in each spatial bin using {\tt THESTORM} \citep[][]{bernard2015letter, bernard2018MNRAS}. {\tt THESTORM} finds the best selection among synthetic stars (divided into simple stellar populations of given ages and metallicities) to mimick the observed distribution of stars in specific regions of the CMD \citep[`bundles';][]{2010ApJ...720.1225M, 2020arXiv200209714R}. We encourage the reader to review \citet[][]{ruizlara2018} for an evaluation of the robustness of {\tt THESTORM} regarding the treatment of binaries or bundle strategy. 

An important factor in obtaining reliable SFHs is that the whole main sequence, down to the oldest main sequence turn-off (around M$_{i}$~=~3.0 for the LMC), is well-represented in the CMD fitting process. For this reason, and also to take into account spatial variations of the crowding levels within the LMC, the faintest magnitude of stars included in the fit was varied accordingly. This value is set to M$_{i}$~=~4.5 if the 20\% completeness level is fainter than M$_{i}$~=~4.5, otherwise the 20$\%$ completeness magnitude is used. For the current study, spatial bins that are highly affected by crowding (50$\%$ completeness magnitude brighter than M$_{i}$~=~3, mainly concentrated in the bar region) are masked out. For the reconstruction of the 2D stellar content of the LMC, we assigned to each HEALPix pixel within the LMC the SFH corresponding to its spatial bin. More detailed information on the process of SFH determination can be found in \citet[][]{2010ApJ...720.1225M}, \citet[][]{ruizlara2018}  or \citet[][]{2020arXiv200209714R}; see also \citet[][]{2011ApJ...730...14H} for a computation of uncertainties; and Ruiz-Lara et al. (in prep.), for the particular case of SMASH data. 

\section{Results}
\label{results}

\begin{figure*}
\centering
\includegraphics[width=0.91\textwidth]{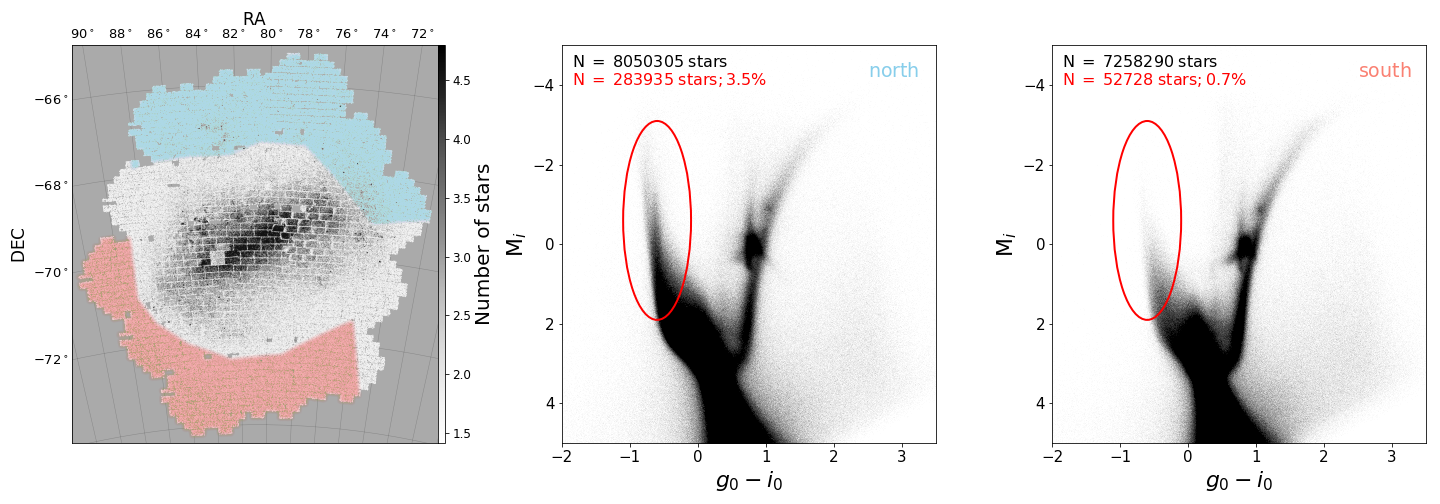}
\caption{Stellar spatial distribution and observed colour-magnitude diagrams corresponding to two visually selected regions representative of the northwest (blue) and southeast (red) regions of the LMC. (Left) LMC-reconstructed stellar density image from SMASH data showing the number density of stars brighter than M$_{i}$~=~-1. North is up, East is left. (Middle) Northwest region, roughly corresponding with the arm. (Right) Southeast region. We note the clear differences in the bright, blue main sequence (approximately delimited by the red ellipse), indicating the presence of a prominent young stellar population in the arm region ($\sim$3.5$\%$ of the 8~million stars sampled against the only $\sim$0.7$\%$ of 7.3 million stars in the south region).}
\label{fig:CMD_arm_counterarm}
\end{figure*}

The northwestern region of the LMC disc sparkles with the presence of H${\rm \alpha}$ emission \citep[e.g.][]{2013MNRAS.436..604R}, recent star formation \citep[e.g.][]{2009AJ....138.1243H}, classical cepheids \citep[][]{2014MNRAS.437.2702M}, and young, massive stars \citep[][]{2019MNRAS.490.1076E}. In contrast, the rest of the LMC disc is largely devoid of evidence supporting the presence of stars as young as those found in the north. This north--south difference exists even beyond the high surface brightness regions: \citet[][]{2016ApJ...825...20B}, using deep optical images, resolved the stellar substructures (stellar arcs and multiple spiral arms) discovered by \citet[][outside the area covered in this work]{1972VA.....14..163D} and confirmed that there is no counterpart to these structures in the south \citep[see also][]{2018ApJ...858L..21M}. This dichotomy is apparent in the SMASH dataset as well \citep[see Fig.\ 2 in][]{2018ApJ...869..125C}. 

The middle and right-hand panels of Fig.~\ref{fig:CMD_arm_counterarm} show the characteristic CMD of stars in the spiral arm (north, blue) and those representative of an approximately symmetric area south of the bar (south, red), respectively (visually selected). Whereas a clear and prominent young main sequence is evident in the northern region, it is virtually absent from the southern region. It is clear that the LMC spiral arm presents contrasting properties with the rest of the LMC disc. In the following, and through the analysis of the SFH of the two selected regions, we seek observational evidence regarding the stability of this spiral structure, its longevity, and the possible mechanisms behind its formation and survival.

\begin{figure}
\centering
\includegraphics[width=0.45\textwidth]{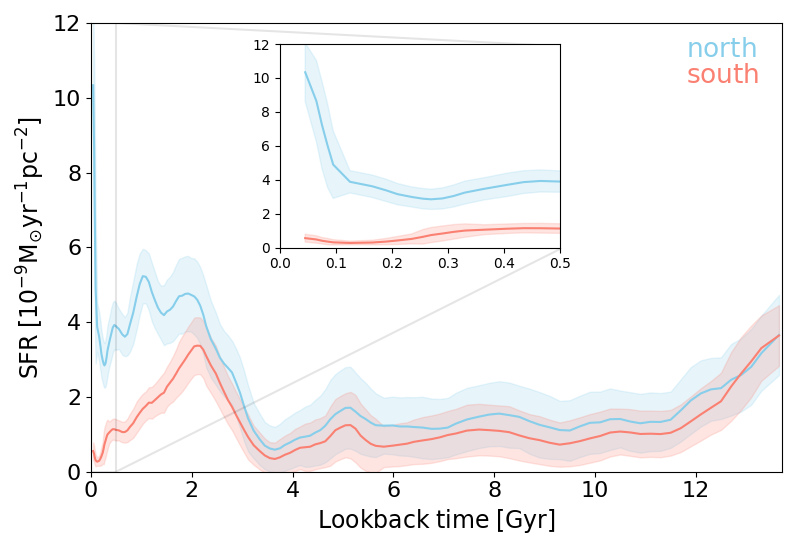}
\caption{Average star formation histories characteristic of the northwest (blue, arm) and southeast (red) regions of the LMC disc. A magnified inset highlights the youngest burst. Shaded regions represent uncertainties in the SFH recovery computed as described in \citet[][]{hidalgo2011}.}
\label{fig:north_south_SFH}
\end{figure}

Figure~\ref{fig:north_south_SFH} compares the SFHs of these two opposed regions. These SFHs are computed by averaging the SFHs for the HEALPix pixels within the defining regions with suitable normalisation\footnote{the normalisation is given by the number of stars brighter than M$_{i}$~=~0.7 in each HEALPix} (left panel of Fig.~\ref{fig:CMD_arm_counterarm}). The similarities of both SFHs at early times until $\sim$2.3~Gyr ago is astonishing, suggesting an efficient mixing of the old stellar populations or a homogeneity of the LMC SFH at early times. Both regions display an important old population (older than 11.5 Gyr), followed by a residual (non-negligible) star formation until $\sim$3~Gyr ago, when the SFR increased significantly. In the last $\sim$2.3 Gyr, the northern region (arm) kept forming stars at a high rate, whereas the southern region progressively decreased its star formation until it reached its current low level. We also find clear secondary bursts at 2, 1, 0.5, and 0.1~Gyr ago in the arm region with hints in the southern region as well. These secondary bursts coincide with those found by \citet[][]{2009AJ....138.1243H}, with the exception of the enhancement at 1 Gyr \citep[detected also in][]{2014MNRAS.438.1067M, 2018MNRAS.473L..16M}. The higher age resolution of our current study allows us to date the re-ignition of star formation roughly 3~Gyr ago with better precision than previous studies~\citep{2009AJ....138.1243H, 2013MNRAS.431..364W}, coinciding with the values found in the bar and inner disc \citep[][]{2018MNRAS.473L..16M}, as well as in the outer LMC disc \citep[][]{2014MNRAS.438.1067M}. Additionally, earlier phases of subdued star formation have occurred in both regions $\sim$5 and 8 Gyr ago (in overall agreement with \citealt[][]{2014MNRAS.438.1067M} and \citealt[][]{2018MNRAS.473L..16M}).

\begin{figure*}
\centering
\includegraphics[width=0.91\textwidth]{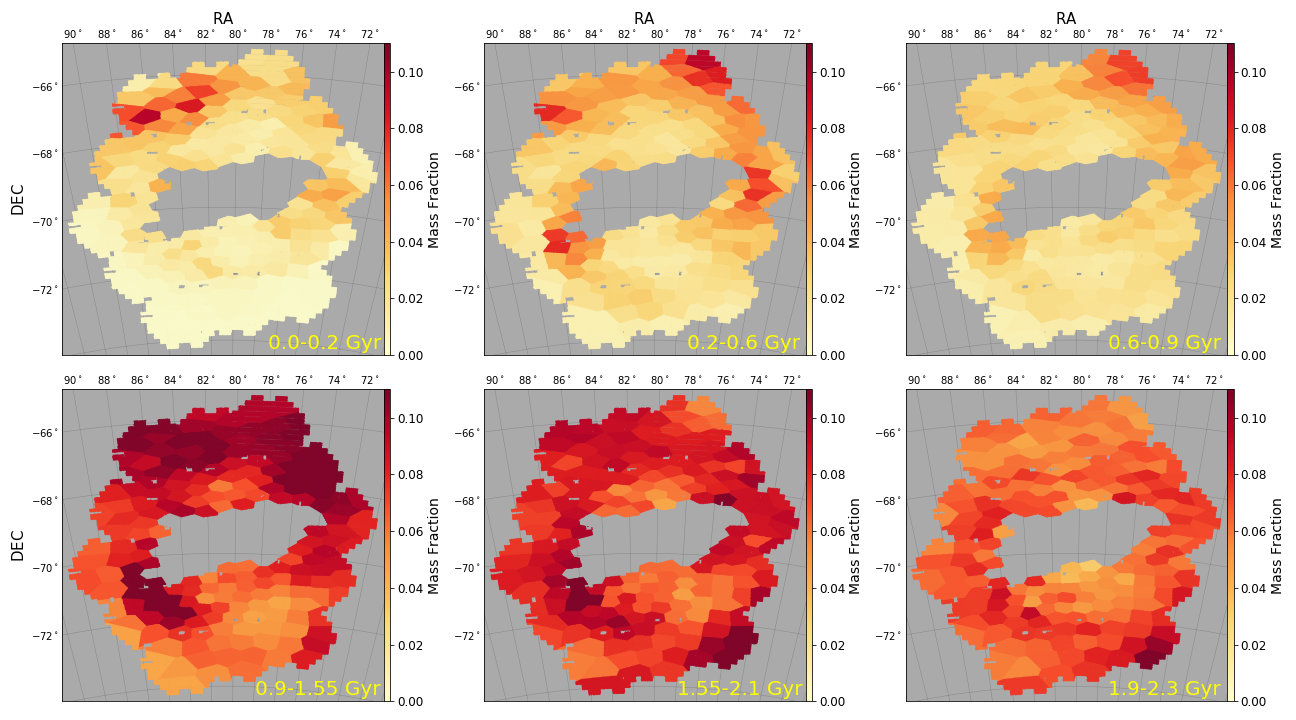}
\caption{Spatial distribution of the stellar mass fractions of different stellar populations (younger than 2.3 Gyr) in the LMC disc. The stellar populations are defined based on their age (indicated in the lower-right corner of each panel); youngest to oldest from the top-left to the bottom-right panels. The mass fraction is normalised to the mass in each spatial bin, i.e. the sum of the fractions of all populations (including older than 2.3~Gyr) in each spatial bin is 1.}
\label{fig:9_mass_fract_map}
\end{figure*}

The combination of large spatial coverage and deep photometry allowed us to derive a SFH with unprecedented high age and spatial resolution in the LMC. As hinted by Fig.~\ref{fig:north_south_SFH}, the distribution of old stars in the LMC lacks clear spatial patterns that could be associated with any LMC morphological component \citep[in agreement with the uniform distribution of RR Lyrae stars presented in][]{2009AcA....59....1S}. However, this changes in the last $\sim$2~Gyr. Figure~\ref{fig:9_mass_fract_map} shows stellar mass fraction maps of stars younger than 2.3~Gyr, divided into age bins defined on the basis of the SFH shown in Fig.~\ref{fig:north_south_SFH}. Whereas stars in the 1.9--2.3~Gyr age range still show no strong spatial coherence, it is for stars younger than 1.55~Gyr (even 2.1~Gyr) that the whole LMC spiral structure --- including the southeastern patch near the end of the bar --- clearly stands out. This structure not only appears to be stable and displays a clear spatial coherence during the last $\sim$2~Gyr, it also shows spatial evolution \citep[see][]{2019MNRAS.490.1076E}. Stars in the 0.6--0.9~Gyr age range accumulate in the northwestern extreme of the arm, whereas stars ranging from 0.2 to 0.6~Gyr old are distributed along the entire arm and the eastern end where the youngest stars are concentrated \citep[the Constellation~III region,][]{2009AJ....138.1243H}.

\section{Discussion and conclusions}
\label{disc_conc}

Morphological asymmetries such as this single-armed structure have been widely attributed to tidal interactions, dwarf-dwarf mergers, and cosmological gas accretion \citep[e.g.][]{1994AJ....107.1320O, 1997ApJ...477..118Z, 2002ApJ...580..705K, 2005A&A...438..507B, 2012MNRAS.421.2109B, 2014MNRAS.439.1948Y, 2016ApJ...827..149P}. Although we cannot rule out some contribution from the latter, the similar enrichment histories that we find in our SFH determination from SMASH data between the northern and southern regions, and especially the lack of any excess of metal-poor stars in the LMC spiral arm, argue against the infall of pristine gas as a primary driver of this structure. In addition, the fact that the LMC appears to be on its first infall into the MW \citep[][]{2007ApJ...668..949B, 2020ApJ...893..121P} allows us to rule out interactions with the MW as the origin of the spiral structure. Simulations suggest that the earliest the MW could have affected the LMC was via ram pressure stripping (interaction with the circumgalactic medium) $\sim$1~Gyr ago, which would have mainly affected the outermost LMC \citep[][]{2015ApJ...815...77S}. The results reported in this work clearly suggest that the LMC spiral structure has been in place for at least 2~Gyr, well before the LMC started feeling the effects of the MW. It is possible that the LMC-SMC interactions triggered the formation of the LMC spiral structure.

For decades, evidence has existed supporting a LMC-SMC common evolution for several billion years \citep[e.g.][]{1974ApJ...190..291M, 2013MNRAS.431..364W}. In fact, mutual interactions between the Clouds have long been invoked to explain large scale structures such as the Magellanic Bridge, Stream, and Leading Arm, as well as its off-centred bar and single spiral arm \citep[e.g.][]{2007ApJ...668..949B, 2012MNRAS.421.2109B, 2012ApJ...750...36D, 2013ApJ...768..109N, 2015MNRAS.452.4222N}. In this line, an analysis of the kinematic information of the \hi~gas, \citet[][]{2008ApJ...679..432N} traced back part of the Leading Arm and Stream to a LMC origin, dating it to be $\sim$1.74 Gyr old. On the other hand, \citet[][]{2018ApJ...854..142F} analysed spectroscopic data from four AGN sight-lines passing through the Leading Arm to provide compelling evidence that a good fraction of the Leading Arm has its origin in the SMC, dating it back $\sim$2~Gyr ago. In either case, these works suggest that the Leading Arm was generated at the same time as the Stream, following a LMC-SMC close encounter some 1.5 to 2.5~Gyr ago \citep[][]{2018ApJ...854..142F}. Indeed, independent simulations from \citet[][]{2012MNRAS.421.2109B} suggest pericentric passages between the Clouds $\sim$2.7 and 1.1 Gyr ago as well as a direct collision $\sim$100-300~Myr ago (see also \citealt{2018ApJ...864...55Z} or \citealt{2019A&A...631A..98M}), partly responsible for the LMC peculiar morphology as well as the Bridge \citep[e.g.][]{1985Natur.318..160I}. Some recent studies also suggest that LMC-SMC interactions are behind newly-discovered stellar structures such as the ring-like stellar overdensity and the outer warp in the LMC \citep[][]{2018ApJ...866...90C, 2018ApJ...869..125C}. The timing of the proposed encounters generally coincide with star formation enhancements reported in this analysis (see Fig.~\ref{fig:north_south_SFH}).

It is intriguing to see that all recent encounters appear to have triggered star formation in the same spatial region, that is, the spiral arm (currently in the northwest, see Fig.~\ref{fig:9_mass_fract_map}). In fact, the \hi~distribution in the LMC appears globally displaced towards this direction, offset with respect to the LMC bar \citep[][]{2003MNRAS.339...87S}. Our results would suggest that this lopsidedness of the gas may have been maintained for the last 2--3 Gyr, with later interactions inducing stochastic self-propagating star formation \citep[][]{1980ApJ...242..517G} where \hi~resides. As asymmetries induced by tidal encounters seem to last several Gyrs \citep[][]{1998ApJ...496L..13L, 2001MNRAS.328.1064N, 2003MNRAS.341..343B, 2016ApJ...825...20B, 2016ApJ...827..149P}, this would explain the notable spatial coherence in the spiral arm stellar populations for the last $\sim$2~Gyr and further support a link between this structure and LMC-SMC interactions dating back $\sim$2--3~Gyr. Theoretical predictions focused on the SFH of the Clouds are needed to confirm this scenario and put the LMC spiral structure within the framework of the dynamical spirals present in more massive systems \citep[][]{1964ApJ...140..646L, 1966ApJ...146..810J}.

\vspace{5mm}

In this work, we present an updated analysis of the SFH over a large area of the outer disc of the LMC with an unparalleled combination of spatial coverage and time resolution. For this purpose, we combine the deep SMASH photometric data with the powerful approach of CMD fitting to study SFHs in galaxies. We find that the spiral arm exhibits a distinctly young population (younger than $\sim$2~Gyr). This spatial coherence provides indisputable evidence that this structure has been in place for at least the last $\sim$2~Gyr, surviving dynamical arguments such as differential rotation. The dating of the spiral arm, together with the orbital history of the Clouds --- in the first infall towards the MW --- strongly advocate for LMC-SMC interactions during the last 3~Gyr as the origin of this stable structure. These results provide key constraints for all future models of the Magellanic System, which should also explain the recent SFH exhibited by the northern spiral arm.

\begin{acknowledgements}

The authors are grateful to the anonymous referee for their invaluable comments improving the original manuscript. TRL, CG, MM, FS and LM acknowledge financial support through grants (AEI/FEDER, UE) AYA2017-89076-P, AYA2016-77237-C3-1-P (RAVET project) and AYA2015-63810-P, as well as by Ministerio de Ciencia, Innovaci\'on y Universidades (MCIU), through Juan de la Cierva - Formaci\'on grant (FJCI-2016-30342) and the State Budget and by Consejer\'\i a de Econom\'\i a, Industria, Comercio y Conocimiento of the Canary Islands Autonomous Community, through Regional Budget. TRL also acknowledges support from the Spanish Public State Employment Service (SEPE). Y.C. acknowledges support from NSF grant AST 1655677. SC acknowledges support from Premiale INAF ``MITIC'' and grant AYA2013-42781P from the Ministry of Economy and Competitiveness of Spain, he has also been supported by INFN (Iniziativa specifica TAsP). M-RC acknowledges support from the European Research Council (ERC) under the European Union's Horizon 2020 research and innovation program (grant agreement No. 682115). SRM acknowledges support through NSF grant AST-1909497. DMD acknowledges support through grant AYA2016-81065-C2-2 and through the ``Centre of Excellence Severo Ochoa'' award for the Instituto de Astrof\'isica de Andaluc\'ia (SEV-2017-0709) and from grant PGC2018-095049-B-C21. AM acknowledges financial support from FONDECYT Regular 1181797 and funding from the Max Planck Society through a Partner Group grant. R.R.M. acknowledges partial support from project BASAL AFB-$170002$ as well as FONDECYT project N$^{\circ}1170364$. This research makes use of python (\url{http://www.python.org}); Matplotlib \citep[][]{hunter2007} and Astropy \citep[][]{astropy2013, 2018AJ....156..123A}.

Based on observations at Cerro Tololo Inter-American Observatory, National Optical Astronomy Observatory (NOAO Prop. ID: 2013A-0411 and 2013B-0440; PI: Nidever), which is operated by the Association of Universities for Research in Astronomy (AURA) under a cooperative agreement with the National Science Foundation. This project used data obtained with the Dark Energy Camera (DECam), which was constructed by the Dark Energy Survey (DES) collaboration. Funding providers for the DES Projects can be found in \url{https://www.darkenergysurvey.org/collaboration-and-sponsors/}.

\end{acknowledgements}

%
%

\bibliographystyle{aa}
\bibliography{bibliography_LMC}


\end{document}